\documentclass[letterpaper,10 pt,conference,dvipsnames]{ieeeconf}
\IEEEoverridecommandlockouts

\let\labelindent\relax

\usepackage{geometry}
\usepackage[absolute,overlay]{textpos}
\usepackage{times}
\usepackage{enumitem}

\usepackage[usenames,dvipsnames]{xcolor}

\usepackage{amsmath,amssymb,mathtools,nccmath}

\usepackage{amsthm}
\usepackage{bbold,dsfont,bbm}

\usepackage{graphicx}
\usepackage{tikz,tikz-cd,pgfplots}
\pgfplotsset{compat=1.17}

\usepackage{hyperref}
\usepackage[capitalise,noabbrev]{cleveref}
\usepackage{cite}

\usepackage{multirow}
\usepackage{booktabs}
\usepackage[acronym]{glossaries}

\usepackage{comment}

\usepackage{units}
\usepackage[per-mode=symbol]{siunitx}
\DeclareSIUnit{\eur}{\euro}
\DeclareSIUnit{\usd}{USD}
\DeclareSIUnit{\mph}{mph}
\DeclareSIUnit{\month}{month}
\DeclareSIUnit{\year}{year}
\DeclareSIUnit{\million}{Mil}
\DeclareSIUnit{\mile}{mile}
\DeclareSIUnit{\car}{car}
\DeclareSIUnit{\train}{train}
\DeclareSIUnit{\mmveh}{\text{$\mu$}MV}
\DeclareSIUnit{\nounit}{-}
\usepackage[font=footnotesize]{caption}
\usepackage{subcaption}

\definecolor{lightblue}{rgb}{0.60784,0.76078,0.90196}
\definecolor{darkblue}{rgb}{0.26667,0.44706,0.76863}
\definecolor{lightgreen}{rgb}{0.66275,0.81569,0.55686}
\definecolor{darkgreen}{rgb}{0.43922,0.67843,0.27843}
\definecolor{orange}{rgb}{0.92941,0.49020,0.19216}
\definecolor{yellow}{rgb}{1.00000,0.75294,0.00000}
\definecolor{grey}{rgb}{0.64706,0.64706,0.64706}
\definecolor{purple}{rgb}{0.51373,0.23529,0.04706}
\newacronym{abk:amod}{AMoD}{Autonomous Mobility-on-Demand}
\newacronym{abk:iamod}{\mbox{I-AMoD}}{intermodal \gls{abk:amod}}
\newacronym{abk:av}{\mbox{AV}}{autonomous vehicle}
\newacronym{abk:sv}{\mbox{SV}}{standard vehicle}

\newacronym{abk:br}{BR}{Best Response}
\newacronym{abk:bpr}{BPR}{Bureau of Public Roads}
\newacronym{abk:bev}{BEV}{Battery Electric Vehicle}
\newacronym{abk:hev}{HEV}{Hybrid Electric Vehicle}
\newacronym{abk:ca}{CA}{congestion-aware}
\newacronym{abk:cara}{CARS}{congestion-aware routing scheme}
\newacronym{abk:cpo}{CPO}{complete partial order}
\newacronym{abk:cdp}{CDP}{co-design problem}
\newacronym{abk:cdpi}{CDPI}{co-design problem with implementation}
\newacronym{abk:dp}{DP}{design problem}
\newacronym{abk:dpi}{DPI}{design problem with implementation}
\newacronym{abk:dcpo}{DCPO}{directed complete partial order}
\newacronym{abk:es}{ES}{e-scooter}
\newacronym{abk:ffcs}{FFCS}{free floating car sharing systems}
\newacronym{abk:ghg}{GHG}{greenhouse gas}
\newacronym{abk:icev}{ICEV}{Internal Combustion Engine Vehicle}
\newacronym{abk:kpi}{KPIs}{Key Performance Indicators}
\newacronym{abk:lw}{LW}{Lightweight}
\newacronym{abk:mm}{{$\mu$}M}{micromobility}
\newacronym{abk:mod}{MoD}{Mobility-on-Demand}
\newacronym{abk:msp}{MSP}{Mobility Service Provider}
\newacronym{abk:mcdp}{MCDP}{Monotone Co-Design Problem}
\newacronym{abk:mcfp}{MCFP}{multi-commodity flow problem}
\newacronym[plural=NE,firstplural=Nash Equilibria (NE)]{abk:ne}{NE}{Nash Equilibrium}
\newacronym{abk:nyc}{NYC}{New York City}
\newacronym{abk:poset}{poset}{partially ordered set}
\newacronym{abk:sb}{SB}{shared bike}
\newacronym{abk:spp}{SPP}{shortest path problem}
\newacronym{abk:kdspp}{k-dSPP}{k-disjoint \gls{abk:spp}}
\newacronym{abk:su}{SU}{Sport Utility}
\newacronym{abk:SE}{SE}{Stackelberg Equilibrium}
\newacronym{abk:sdp}{SDP}{short-distance price}
\newacronym{abk:ldp}{LDP}{long-distance price}
\newcommand{\destination}{d}

\newcommand{\delay}[1]{\ell_{#1}}
\newcommand{\delayTime}[1]{\bar\ell_{#1}}

\newcommand{\delayTimeInverse}[1]{\bar\ell_{#1}^{-1}}

\newcommand{\origin}{o}
\newcommand{\action}[1]{u_{#1}}

\newcommand{\setOfArcs}{\mathcal{A}}

\newcommand{\graph}{\mathcal{G}}

\newcommand{\mat}[1]{\mathbf{#1}}

\newcommand{\pureStrategy}[1]{\gamma_{#1}}

\newcommand{\tup}[1]{\left\langle #1 \right\rangle}
\newcommand{\actionSet}[1]{\mathcal{U}_{#1}}
\newcommand{\utility}[1]{U_{#1}}
\newcommand{\valueTime}{V_\mathrm{T}}
\newcommand{\setOfVertices}{\mathcal{V}}

\newcommand{\pureStrategySet}[1]{\Gamma_{#1}}
\newcommand{\rate}{\alpha}

\newcommand{\true}[1]{\texttt{T}}

\newcommand{\derivative}[3]{\left.\frac{\mathrm{d}#1}{\mathrm{d}#2}\right|_{#3}}
\newcommand{\secondDerivative}[3]{\left.\frac{\mathrm{d}^2#1}{\mathrm{d}#2^2}\right|_{#3}}
\newcommand{\proj}[2]{\Pi_{#2}\left[#1\right]}
\renewcommand{\baselinestretch}{0.91}
\theoremstyle{definition}
\newtheorem{assumption}{Assumption}
\newtheorem*{assumption*}{Assumption}
\newtheorem{theorem}{Theorem}

\newtheorem{lemma}[theorem]{Lemma}
\newtheorem{corollary}[theorem]{Corollary}

\newtheorem{proposition}[theorem]{Proposition}
\newtheorem{definition}{Definition}
\theoremstyle{remark}

\Crefname{figure}{Fig.}{Figures}
\makeatletter
    \if@cref@capitalise
        \crefname{subsection}{Section}{Sections}
        \crefname{subsubsection}{Section}{Sections}
        \crefname{assumption}{Assumption}{Assumptions}
        \crefname{problem}{Problem}{Problems}
    \else
        \crefname{subsection}{section}{sections}
        \crefname{subsubsection}{section}{sections}
        \crefname{assumption}{assumption}{assumptions}
        \crefname{problem}{problem}{problems}
    \fi
\makeatother

\usetikzlibrary{positioning,intersections,hobby,patterns,calc,decorations.pathmorphing,decorations.markings, shadows,shapes}
\tikzstyle{block} = [draw, rectangle, minimum height=2em, minimum width=3em,font=\bfseries,rounded corners,thick]
\tikzstyle{block1} = [draw, rectangle, minimum height=1.5em, minimum width=2.5em]
\tikzstyle{blockDyn} = [draw, rectangle, minimum height=2.5em, minimum width=3.5em, align=center, inner sep=10pt, thick, fill=white, copy shadow={draw=black,fill=black,opacity=1,shadow xshift=0.5ex,shadow yshift=-0.5ex}]
\tikzstyle{blockAlg} = [draw, rectangle, minimum height=1.5em, minimum width=2.5em, align=center, inner sep=10pt, thick]
\tikzstyle{sum} = [draw,circle]

\tikzstyle{nodePre} = [circle, draw,inner sep=1pt,node contents={$\preceq$},thick]
\tikzstyle{nodePreEmpty} = [circle, draw,inner sep=1pt,thick]
\tikzstyle{nodePos} = [circle, draw,inner sep=1pt,node contents={$\posceq$},thick]
\tikzstyle{nodeProd} = [rectangle, draw,inner sep=4pt,node contents={$\times$},rounded corners,thick]
\tikzstyle{nodeSum} = [rectangle, draw,inner sep=4pt,node contents={$\mathbf{+}$},rounded corners,thick]

\definecolor{DPgreen}{rgb}{0.0, 0.5, 0.0}
\definecolor{red}{rgb}{0.75, 0.0, 0.0}

\newif\ifmargincomments %
\margincommentstrue

\newif\ifextendedversion %

\ifmargincomments
	
	\newcommand{\nlmargin}[2]{{\color{blue}#1}\marginpar{\color{blue}\raggedright\footnotesize [NL]:#2}}
	\newcommand{\gzmargin}[2]{{\color{red}#1}\marginpar{\color{red}\raggedright\footnotesize [GZ]:#2}}
\else
	
	\newcommand{\nlmargin}[2]{#1}
	\newcommand{\gzmargin}[2]{#1}
\fi

\title{
\textbf{Game Theory to Study Interactions between Mobility Stakeholders}
}
\author{Gioele Zardini$^{1}$, Nicolas Lanzetti$^{2}$, Laura Guerrini$^{2}$, Emilio Frazzoli$^{1}$, Florian Dörfler$^{2}$
\thanks{
$^{1}$Institute for Dynamic Systems and Control, ETH Z\"urich, Switzerland {\tt \{gzardini,efrazzoli\}@ethz.ch}.}
\thanks{
$^{2}$Automatic Control Laboratory, ETH Z\"urich, Switzerland, {\tt \{lnicolas,laurague,dorfler\}@ethz.ch}.}
\thanks{
The first two authors contributed equally to this work.
This work was supported by the Swiss National Science Foundation under NCCR Automation, grant agreement 51NF40\_180545.}
}

\begin{document}

\begin{textblock*}{\textwidth}(15mm,18mm) %
\bf \textcolor{NavyBlue}{Proceedings of the 2021 IEEE International Conference on Intelligent Transportation Systems (Best Paper Award)}
\end{textblock*}

\maketitle
\begin{abstract}
Increasing urbanization and exacerbation of sustainability goals threaten the operational efficiency of current transportation systems and confront cities with complex choices with huge impact on future generations. At the same time, the rise of private, profit-maximizing \acrlongpl{abk:msp} leveraging public resources, such as ride-hailing companies, entangles current regulation schemes.
This calls for tools to study such complex socio-technical problems.
In this paper, we provide a game-theoretic framework to study interactions between stakeholders of the mobility ecosystem, modeling regulatory aspects such as taxes and public transport prices, as well as operational matters for \acrlongpl{abk:msp} such as pricing strategy, fleet sizing, and vehicle design.
Our framework is modular and can readily accommodate different types of \acrlongpl{abk:msp}, actions of municipalities, and low-level models of customers' choices in the mobility system. 
Through both an analytical and a numerical case study for the city of Berlin, Germany, we showcase the ability of our framework to compute equilibria of the problem, to study fundamental tradeoffs, and to inform stakeholders and policy makers on the effects of interventions. Among others, we show tradeoffs between customers' satisfaction, environmental impact, and public revenue, as well as the impact of strategic decisions on these metrics. 
\end{abstract}

\section{Introduction}
\label{sec:introduction}
In past decades, cities worldwide have observed a dramatic urbanization. Today, \SI{55}{\percent} of the world's population resides in urban areas, and by 2050 the proportion is expected to reach \SI{68}{\percent}~\cite{un2020}. A direct consequence of the population density growth is the increase of urban travel, and of the externalities it produces~\cite{czepkiewicz2018urbanites}. In this rapidly expanding setting, cities have to take important decisions to adapt their transportation system to welcome larger travel demands. This is a very complex task for at least three reasons. 
First, cities need to accommodate the changing travel needs of the population, by predicting them~\cite{calastri2019people}, and by ensuring fairness and equity~\cite{ranchordas2020smart}. 
Second, designed policies not only have to account for the citizens' satisfaction, but also for their impact on private \glspl{abk:msp} such as ride-hailing companies, \gls{abk:mm}, and, in a near future, \gls{abk:amod} systems~\cite{zardinilanzettiAR2021}. Indeed, such services gained a considerable share of the transportation market in recent years; e.g., in NYC, ride-hailing companies have increased their daily trips by \SI{1000}{\percent} from 2012 to 2019~\cite{OneNYC}. While offering more choices to travellers, these systems operate benefiting from public resources (such as roads and public spaces), are profit-oriented, and often lead to potentially disruptive consequences for the efficiency of the transportation system and for society at large~\cite{berger2018drivers,rogers2015social,yigitcanlar2019disruptive}. In this avenue, cities gain an important, onerous regulatory role.
Third, policies have to be designed while meeting global sustainability goals. It is not surprising that cities are estimated to be responsible for \SI{78}{\percent} of the world's energy consumption and for over \SI{60}{\percent} of the global greenhouse emissions (\SI{30}{\percent} of which is produced by transportation, in US)~\cite{un2020bis}. Indeed, sustainability is central in policy-making worldwide: NYC plans to increase sustainable trips from \SI{68}{\percent} to \SI{80}{\percent}~\cite{OneNYC}, and EU plans a \SI{90}{\percent} reduction of emissions by 2050~\cite{euplan2020}.

Taken together, the aforementioned perspectives highlight the complexity of this socio-technical problem, and imperatively call for methods to inform and drive policy makers. 

The goal of this paper is to lay the foundations for a framework through which one can model sequential, competitive interactions between stakeholders of the mobility ecosystem, and characterize their equilibria. Specifically, we leverage game theory to frame the problem in a modular fashion, and provide both an analytical and a numerical case study to showcase our methodology. 

\subsection{Related Literature}
Our work lies at the interface of applications of game-theory in transportation science, and policy making related to future mobility systems. Game theory has been leveraged to solve various mobility-related problems. %
Main application areas include optimization of pricing strategies for \glspl{abk:msp}~\cite{mingbao2010pricing, gong2014analysis, yang2019subsidy,chen2016management,kuiteing2017network, bimpikis2019spatial}, analysis of interactions between \gls{abk:msp} and users~\cite{dandl2019autonomous,turan2020competition,lei2018evolutionary}, interactions between authorities and \glspl{abk:msp}~\cite{di2019unified,hernandez2018game,mo2021dynamic,balac2019modeling,lanzetti2019self, LanzettiSchifferEtAl2021, sun2019evolutionary, wang2019mitigation}, and tolls and incentives to regulate congested networks~\cite{swamy2012effectiveness,paccagnan2021optimal, lazar2020optimal,krichene2017stackelberg,zhou2015optimal,koryagin2018urban,mei2017game,bianco2016game}.
While~\cite{mingbao2010pricing, gong2014analysis} use game theory to determine prices for public transport,~\cite{yang2019subsidy, chen2016management} focus on subsidies and management of shared fleets of electric vehicles, and~\cite{kuiteing2017network,bimpikis2019spatial} focus on pricing strategies at the network level. 
The competition among \glspl{abk:msp} is studied in~\cite{dandl2019autonomous} through a real-time gaming framework, in~\cite{turan2020competition} focusing on~\gls{abk:amod} systems, and via evolutionary game theory with a focus on ride-sourcing in~\cite{lei2018evolutionary}.
When studying interactions between authorities and \glspl{abk:msp}, shared mobility systems received much attention. In particular,~\cite{di2019unified} proposes a unified game-theoretic framework for policy making related to shared mobility systems,~\cite{hernandez2018game} focuses on carpooling systems and evolutionary stable policies,~\cite{mo2021dynamic} analyzes the dynamic interactions between shared \glspl{abk:av} and public transit, and~\cite{balac2019modeling} proposes a modeling framework for competing carsharing operators in Zurich, Switzerland. 
Furthermore, game-theoretic frameworks to study interactions of \glspl{abk:msp} and public transportation systems are proposed in~\cite{lanzetti2019self, LanzettiSchifferEtAl2021} (for \glspl{abk:av}), in~\cite{sun2019evolutionary} (for ride-hailing companies), and in~\cite{wang2019mitigation} (for bike-sharing systems). 
General game-theoretic approaches for the regulation of congestion via policies and incentives have been studied in~\cite{swamy2012effectiveness, paccagnan2021optimal, lazar2020optimal}, and via clever routing and price-making in~\cite{krichene2017stackelberg, zhou2015optimal}. Moreover,~\cite{koryagin2018urban} focuses on game-theoretic urban planning to reduce externalities of transportation systems,~\cite{mei2017game} leverages mobility patterns to propose travel incentives, and~\cite{bianco2016game} studies tolling policies for the transportation of hazardous materials.

There has been research on policy making for future mobility systems not involving game theory \cite{fullerton2002can,iwata2016can,zhang2016optimal, slowik2019can, chremos2020socially, zardini2020co,zoepf2018economics,ostrovsky2019carpooling}. Strategies to reduce externalities (including tolls, subsidies, electrification) are proposed in~\cite{fullerton2002can,iwata2016can,zhang2016optimal,slowik2019can}, and socially efficient arguments are made in~\cite{chremos2020socially,zardini2020co}. 
Finally the economics of ride-hailing, \glspl{abk:av}, and carpooling companies is studied in~\cite{zoepf2018economics,ostrovsky2019carpooling}.

Overall, all these works either focus on specific problems, neglect some mobility stakeholders, or ignore game-theoretic dynamics. So, to the best of our knowledge, there does not exist a comprehensive framework which allows one to formulate and solve mobility problems involving interactions between different stakeholders, at different time-scales, all the way from municipalities to customers, through \glspl{abk:msp}.

\subsection{Statement of Contribution}
To fill this gap, we study the interactions between a central municipality, \glspl{abk:msp}, and customers from a game-theoretic perspective. Our contribution is threefold. 
First, we provide a general game-theoretic framework to model the sequential and simultaneous interactions between a municipality and \glspl{abk:msp}. 
Second, we instantiate our framework with two low-level models of the mobility system: a parallel-arc congestion game and a game-theoretic model including ride-hailing companies. 
Third, we present numerical results for the city of Berlin, and derive insights to inform stakeholders of the mobility ecosystem.

\subsection{Organization}
The remainder of this paper is as follows.
We specify our problem setting and model in \cref{sec:model}. In \cref{sec:results}, we detail our case study and present numerical results. We draw conclusions and present an outlook on future research in \cref{sec:conclusion}. Proofs are relegated to the appendix.

\section{Mobility Interactions as Sequential Games}
\label{sec:model}
As outlined in \cref{sec:introduction}, urban mobility systems feature a broad variety of complex interactions. We classify such interactions based on the time scale at which they occur. We identify four time scales: a day, a month, a year, and five years. While day-to-day interactions include specific operational conditions such as dynamic pricing and rebalancing policies for \glspl{abk:msp}, monthly interactions cover changes in the regions served by a particular \gls{abk:msp}. When looking at a horizon of one year, one can consider general price plans for public transport, taxation systems on ride-hailing companies, as well as logistic decisions for \glspl{abk:msp}, such as fleet sizing and fleet diversification. Finally, a horizon of five years could include infrastructural changes, land use planning, and public contracts for transportation systems. In the following, we detail a model for the yearly horizon. This way we consider long-term horizon interventions (e.g., transportation network topology) as fixed parameters and include short-term horizon dynamics (e.g., mode selection, rebalancing, etc.) in what we call \emph{low-level} model of the mobility system. Nonetheless, our methodology readily applies to the other settings as well. 

\subsection{Game-theoretic Model}
\begin{figure}[t]
    \begin{center}
    \begin{tikzpicture}[scale=0.9]
        \draw[rounded corners,fill=blue!20!white,color=blue!20!white] (-3.4, -1.75) rectangle ++(2.8,0.5) {};
        \draw[rounded corners,fill=blue!20!white,color=blue!20!white] (0.6, -1.75) rectangle ++(2.8,0.5) {};
        \draw[rounded corners,fill=red!20!white,color=red!20!white] (1.7, -1) rectangle ++(0.6,0.5) {};
        \draw[rounded corners,fill=red!20!white,color=red!20!white] (-1.7, -1) rectangle ++(-0.6,0.5) {};
        
        \draw[rounded corners,thick,color=ForestGreen!20!white,fill=ForestGreen!20!white] (-0.3, -0.25) rectangle ++(0.6,0.5) {};
        
        \node[circle,draw,thick] at (0,0) (mun) {};
        \node[circle,font=\footnotesize] at (4,0)  {\textbf{Municipality}};
        
        \node[circle,draw,thick] at (-2,-0.75) (msp11) {}; 
        \node[circle,draw,thick] at (2,-0.75) (msp12) {}; 
        \node[circle,font=\footnotesize] at (4,-0.75)  {\textbf{\gls{abk:msp} 1}};
        
        \node[circle,draw,thick] at (-3,-1.5) (msp21) {};
        \node[circle,draw,thick] at (-1,-1.5) (msp22) {}; 
        \node[circle,draw,thick] at (1,-1.5) (msp23) {}; 
        \node[circle,draw,thick] at (3,-1.5) (msp24) {}; 
        \node[circle,font=\footnotesize] at (4,-1.5)  {\textbf{\gls{abk:msp} 2}};
        
        \tikzstyle{myrectangle} = [rectangle,font=\tiny,align=center,rounded corners,fill=black!10!white,inner sep=1.5pt]
        \node[myrectangle] at (-3.6,-2.5) (o11) {$U_0$ \\ $U_1$ \\ $U_2$}; 
        \node[myrectangle] at (-3.0,-2.5) (o12) {$U_0$ \\ $U_1$ \\ $U_2$}; 
        \node[myrectangle] at (-2.4,-2.5) (o13) {$U_0$ \\ $U_1$ \\ $U_2$}; 
        \node[myrectangle] at (-1.6,-2.5) (o21) {$U_0$ \\ $U_1$ \\ $U_2$}; 
        \node[myrectangle] at (-1.0,-2.5) (o22) {$U_0$ \\ $U_1$ \\ $U_2$}; 
        \node[myrectangle] at (-0.4,-2.5) (o23) {$U_0$ \\ $U_1$ \\ $U_2$}; 
        \node[myrectangle] at (1.6,-2.5) (o31) {$U_0$ \\ $U_1$ \\ $U_2$}; 
        \node[myrectangle] at (1.0,-2.5) (o32) {$U_0$ \\ $U_1$ \\ $U_2$}; 
        \node[myrectangle] at (0.4,-2.5) (o33) {$U_0$ \\ $U_1$ \\ $U_2$}; 
        \node[myrectangle] at (3.6,-2.5) (o41) {$U_0$ \\ $U_1$ \\ $U_2$}; 
        \node[myrectangle] at (3.0,-2.5) (o42) {$U_0$ \\ $U_1$ \\ $U_2$}; 
        \node[myrectangle] at (2.4,-2.5) (o43) {$U_0$ \\ $U_1$ \\ $U_2$}; 
        
        \draw[thick,-] (mun) -- (msp11); 
        \draw[thick,-] (mun) -- (msp12); 
        
        \draw[thick,-] (msp11) -- (msp21); 
        \draw[thick,-] (msp11) -- (msp22); 
        \draw[thick,-] (msp12) -- (msp23); 
        \draw[thick,-] (msp12) -- (msp24); 
        
        \draw[thick,-] (msp21) -- (o11); 
        \draw[thick,-] (msp21) -- (o12); 
        \draw[thick,-] (msp21) -- (o13);
        \draw[thick,-] (msp22) -- (o21); 
        \draw[thick,-] (msp22) -- (o22); 
        \draw[thick,-] (msp22) -- (o23); 
        \draw[thick,-] (msp23) -- (o31); 
        \draw[thick,-] (msp23) -- (o32); 
        \draw[thick,-] (msp23) -- (o33); 
        \draw[thick,-] (msp24) -- (o41); 
        \draw[thick,-] (msp24) -- (o42); 
        \draw[thick,-] (msp24) -- (o43); 
    \end{tikzpicture}
    \end{center}
    \caption{Sequential interactions of the game in the case of a municipality with two available actions and two \glspl{abk:msp} having two and three available actions, respectively. First, the municipality chooses its action. Then, \glspl{abk:msp} simultaneously decide on their action. The payoff of all stakeholders follows accordingly. The boxes depict the so-called information sets: \gls{abk:msp} 1 knows the action of the municipality, but does not know the action of \gls{abk:msp} 2.}
    \label{fig:sequence}
    \vspace{-0.6cm}
\end{figure}
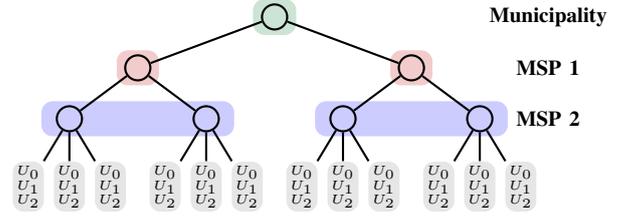

We consider a one-stage sequential game between a municipality and~$N\in\mathbb{N}$ \glspl{abk:msp}, also called single-leader multi-follower Stackelberg game. Herein, the municipality first decides on policies such as taxes, public transport prices, and number of released vehicles licenses, to maximize social welfare. The profit-oriented \glspl{abk:msp} then selfishly co-design their fleet and their pricing strategies (see \cref{fig:sequence}).
Finally, the outcome of the game results from a low-level model of the mobility system, which includes the dynamics happening on a short-term horizon. 

Formally, the game is specified as follows. 
The municipality chooses its action~$\action{0}$ from the (possibly uncountable) set of actions~$\actionSet{0}$. Since the municipality plays first, actions and strategies coincide:~$\pureStrategy{0}=\action{0}$ means that the municipality plays action~$\action{0}\in\actionSet{0}$; so, the set of strategies~$\pureStrategySet{0}$ and the set of actions~$\actionSet{0}$ coincide.
For instance, if the municipality only designs the (flat) price of public transport, then~$\actionSet{0}=\pureStrategySet{0}\coloneqq\mathbb{R}_{\geq 0}$. 
\gls{abk:msp}~$j\in\{1,\ldots,N\}$ selects their action~$\action{j}$ from the (possibly uncountable) set of actions~$\actionSet{j}(\pureStrategy{0})$, possibly dependent on the strategy of the municipality. Since \glspl{abk:msp} play after the municipality, the strategy~$\pureStrategy{j}$ is a map~$\pureStrategy{j}: \pureStrategySet{0}\to\bigcup_{\pureStrategy{0}\in\pureStrategySet{0}}\actionSet{j}(\pureStrategy{0})$,
where~$\pureStrategy{j}(\pureStrategy{0})\in\actionSet{j}(\pureStrategy{0})$.
We denote by~$\pureStrategySet{j}$ the set of all such maps.
For instance,~$\pureStrategy{j}(\pureStrategy{0})\in\actionSet{j}$ is the action that \gls{abk:msp}~$j$ plays if the municipality played~$\pureStrategy{0}\in\pureStrategySet{0}$.
For simplicity, we neglect the influence of~\gls{abk:msp}~$j$ to the set of actions of \gls{abk:msp}~$k\neq j$. Yet, our framework can be readily extended to accommodate such interactions.
Finally, the payoffs of all agents result from a low-level model of the mobility system comprising, among others, day-to-day behavior of the customers, and dynamic pricing of \glspl{abk:msp}. While numerical results very much depend on this model, the latter can be easily replaced, making our framework modular. Formally, we associate to the municipality ($j=0$) and to each \gls{abk:msp}~$j\in\{1,\ldots,N\}$ a payoff function
\begin{equation*}
\begin{aligned}
    \utility{j}:
    \pureStrategySet{0}\times\pureStrategySet{1}\times\ldots\times\pureStrategySet{N}
    &\to
    \mathbb{R} \\
    \tup{\pureStrategy{0},\pureStrategy{1},\ldots,\pureStrategy{N}} &\mapsto \utility{j}(\pureStrategy{0},\pureStrategy{1},\ldots,\pureStrategy{N}).
\end{aligned}
\end{equation*}
In our case studies, we will focus on $\utility{j}$ being the profit for \glspl{abk:msp} and $\utility{0}$ being the social welfare for the municipality. Nonetheless, our framework accommodates players with different interests (e.g., return on investment). 

\subsection{Equilibria}
To characterize equilibria of our game, we use the classical notion of pure equilibrium in sequential games. Intuitively, a tuple of strategies is an equilibrium of the game if no agent is willing to unilaterally deviate from its strategy.
Formally:
\begin{definition}[Equilibrium]
\label{def:equilibrium}
    The tuple~$\tup{\pureStrategy{0}^\ast,\pureStrategy{1}^\ast,\ldots,\pureStrategy{N}^\ast}\in\prod_{i\in \{0,\ldots,N\}}\pureStrategySet{i}$ is an \emph{equilibrium} of the game if for all players $j\in\{0,\ldots,N\}$:~$\utility{j}(\pureStrategy{j}^\ast,\pureStrategy{-j}^\ast)
        \geq 
        \utility{j}(\pureStrategy{j},\pureStrategy{-j}^\ast),
        \forall\pureStrategy{j}\in\pureStrategySet{j}$, where the subscript~$-j$ represents all players but~$j$.
\end{definition}
\cref{def:equilibrium} emphasizes why we distinguish between actions and strategies. In particular, \cref{def:equilibrium} would fail if expressed in terms of actions, as it would ignore the sequential nature of the game. Conversely, strategies, defined as maps from the ``current information'' to the set of actions, capture the sequential nature of the game.
As well-known in game theory, equilibria need not exist: one may easily come up with examples of sequential games with no equilibrium. Nonetheless, we will see that, when one fixes a low-level model of the mobility system, it may be possible to study sufficient conditions for the existence of equilibria. 

\subsection{Discussion}
First, we do \emph{not} a priori fix the low-level model of the mobility system, allowing one to choose the instance which best suits a desired analytical setting. Examples of low-level models include congestion games~\cite{paccagnan2021optimal}, mobility simulators~\cite{ziemke2019matsim}, and approaches based on reaction curves, ubiquitous in economics~\cite{train2009discrete}.
Second, we tacitly assumed a market with a fixed number of \glspl{abk:msp}. This assumption is realistic for the yearly time scale of our game and can be relaxed for other time scales. 
Third, we assumed a sequential game with sequence as in \cref{fig:sequence}. Arguably, one could think about \glspl{abk:msp} acting first, making the municipality a \emph{reactive} player. We believe that the proposed sequence well aligns with the yearly time horizon, forcing \glspl{abk:msp} willing to enter the market to follow rules established by a municipality. Nevertheless, our formulation can accommodate permutations in the action sequences.

\subsection{Specifying Low-level Models}
To showcase our framework, we instantiate it with two different low-level models of the mobility system.
First, we consider a simplistic mobility system resulting from decoupled congestion games on parallel-arc networks. 
This allows for a clear exposition of the analyzed dependencies, and a thorough analytical study of equilibria.
Second, we consider a mobility system whereby a \gls{abk:msp} applies dynamic pricing, and can therefore select its pricing strategy in real-time, and a \gls{abk:msp} which strategically decides operational matter, such as fleet sizing and composition. Customers strategically react to minimize their overall travel cost, resulting from fares paid throughout the trip and monetary value of time. 

\subsubsection{Analytical Parallel-arc Congestion Game}
\begin{figure}[t]
    \centering 
    \input{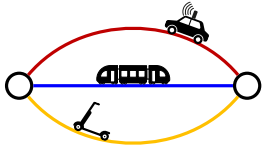}
    \caption{Example of a network with a municipality offering public transport an \gls{abk:amod} system, and a \gls{abk:mm} operator offering mobility service.}
    \label{fig:congestion_game}
    \vspace{-0.5cm}
\end{figure}

We consider one demand (per unit time) between two non-identical nodes, connected by multiple parallel arcs (see \cref{fig:congestion_game}). Each arc denotes an homogeneous mode of transportation, which (possibly combined with walking) leads customers from the origin to the destination node.
The municipality chooses a non-negative price from the compact set~$\actionSet{0}\coloneqq [0,p_0^\mathrm{max}]$, for some~$p_0^\mathrm{max}\in\mathbb{R}_{>0}$, while \glspl{abk:msp} co-design the price of the ride~$p_j$ and their fleet size~$f_j$, i.e.,~$\actionSet{j}\coloneqq\mathbb{R}_{\geq 0}^2$. For simplicity, we assume that \glspl{abk:msp} can choose arbitrarily large fleet sizes and prices. 
To each \gls{abk:msp}~$j\in\{1,\ldots,N\}$ (and therefore to each arc) we assign a non-decreasing smooth delay function
\begin{equation*}
    \delay{j}(\pureStrategy{j}): \mathbb{R}_{\geq 0} \to \mathbb{R}_{\geq 0}\cup\{+\infty\},
\end{equation*}
which captures the total cost of using that arc to reach the destination. We use the notation~$\delay{j}(\pureStrategy{j})$ to emphasize the dependency of the delay function of the strategy of provider~$j$\footnote{Formally,~$\delay{j}$ is a function from the set of strategies to the set of non-negative smooth functions.}. Since customers minimize their individual cost, given as the sum of fares and monetary value of time, we consider delays function of the form  
\begin{equation*}
\begin{aligned}
    \delay{0}(\pureStrategy{0})(z) &= \delay{j}(p_0)(z)
    = p_0 + \valueTime\cdot\delayTime{0}, \\
    \delay{j}(\pureStrategy{j})(z) &= \delay{j}((p_j,f_j))(z)
    =
    \begin{cases}
        p_j + \valueTime\cdot\delayTime{j}(z) & \text{if } z \leq f_j \\
        +\infty & \text{else},
    \end{cases}
\end{aligned}
\end{equation*}
where~$\valueTime\in\mathbb{R}_{>0}$ is the customers' (average) value of time,~$\bar{\ell_0}\in\mathbb{R}_{\geq 0}$ is the total time required to reach the destination with public transport, and~$\delayTime{j}\colon\mathbb{R}_{\geq 0} \to \mathbb{R}_{\geq 0}\cup\{+\infty\}$ is a non-negative non-decreasing smooth function accounting for congestion. For instance, in the case of an \gls{abk:msp} operating on the road network, one may construct~$\delayTime{j}$ based on the BPR function~\cite{united1964traffic}.
In this setting, we model the outcome of the interactions among customers via the well-known notion of Wardrop equilibrium: 
\begin{definition}[Wardrop equilibrium]
$x^\ast\in[0,1]^{N+1}$ with $\sum_{i=0}^Nx_i = 1$ is a \emph{Wardrop equilibrium} if for all $i,j\in\{0,\ldots,N\}$ one has~$\delay{i}(x_i) > \delay{j}(x_j) \Rightarrow x_i=0$.
\end{definition}
In other words, this condition ensures that no agent has an incentive to travel through another arc: at equilibrium an arc can admit a larger delay than other arcs only if no one travels through it. Also, without loss of generality, we assume a total flow of 1. 
It is well-known that congestion games have a Wardrop equilibrium, which can be recovered from the solution of an optimization problem:
\begin{proposition}[Equilibria of congestion games]\label{prop:equilibria_congestion}
The congestion game possesses a Wardrop equilibrium. Moreover, all equilibria coincide with the solutions of
\begin{equation}\label{eq:wardrop_optimization}
\begin{aligned}
    \min_{x\in[0,1]^{N+1}}
    &\sum_{j=0}^N\int_0^{x_j}\delay{j}(z)\,\mathrm{d}z \\
    &\sum_{j=0}^Nx_j = 1, \quad x_j\leq f_j\quad \forall j\in\{1,\ldots,N\}.
\end{aligned}
\end{equation}
Finally, if~$\delay{j}$ for all $j\in\{1,\ldots,N\}$ is strictly increasing, then the Wardrop equilibrium is unique. 
\end{proposition}
To formally talk about \emph{the} equilibrium we make the following assumption: 
\begin{assumption}\label{assumption:delay_1}
The functions $\delayTime{j}$ are convex and strictly increasing for all $j\in\{1,\ldots,N\}$. 
\end{assumption}
\cref{assumption:delay_1} encompasses relevant delay functions, such as the BPR function~\cite{united1964traffic}.
Strict monotonicity, via \cref{prop:equilibria_congestion}, ensures uniqueness of the equilibrium of the congestion game. Hence, we unambiguously denote the Wardrop equilibrium by $x^\ast(\pureStrategy{0},\pureStrategy{1},\ldots,\pureStrategy{N})$.
Armed with this result, we can establish existence of an equilibrium for our game. To do so, we first introduce utilities for all players. 
The socially-aware municipality maximizes social welfare, defined in terms of cost for the customers (including fares and monetary value of time), cost of emissions, and public revenue:
\begin{equation*}
\begin{aligned}
    &\utility{0}(\pureStrategy{0},\pureStrategy{-0})
    \coloneqq
    -k_1\cdot\sum_{j=0}^N \delay{j}(x_j^\ast(\pureStrategy{0},\pureStrategy{-0}))\cdot x_j^\ast(\pureStrategy{0},\pureStrategy{-0}) \\[-9pt]
    &-k_2\cdot\sum_{j=1}^N\varepsilon_j\cdot x_j^\ast(\pureStrategy{0},\pureStrategy{-0})
    +k_3\cdot\pureStrategy{0}\cdot x_0^\ast(\pureStrategy{0},\pureStrategy{-0}),
\end{aligned}
\end{equation*}
where~$k_1$,~$k_2$, and~$k_3$ are strictly positive weights,~$\varepsilon_j$ is the marginal cost of emissions of \gls{abk:msp}~$j$, and~$\pureStrategy{-j}$ collects the strategies of all players but~$j$.
For ease of exposition, we consider profit-maximizing \glspl{abk:msp}:
\begin{equation*}
    \utility{j}(\pureStrategy{j},\pureStrategy{-j})
    \coloneqq
    (p_j - c_j) x_j^\ast(\pureStrategy{j},\pureStrategy{-j}) - \bar c_j f_j,
\end{equation*}
where~$c_j$ denotes the cost of serving a trip and~$\bar c_j$ is the cost of buying a vehicle. We write~$\tilde{c}_j\coloneqq c_j+\bar{c}_j$. To show existence of an equilibrium of the sequential game (c.f. \cref{def:equilibrium}), we rely on the following assumption: 
\begin{assumption}\label{assumption:delay_2}
All $\delayTime{j}$ satisfy
\begin{equation*}
    \sum_{j=1}^N\delayTimeInverse{j}\left(\frac{p_0^\mathrm{max}}{\valueTime}+\delayTime{0}\right)\leq 1.
\end{equation*}
\end{assumption}
\cref{assumption:delay_2} ensures that for all realizations of $\pureStrategy{0}, \pureStrategy{1}, \ldots, \pureStrategy{N}$ at least a portion of the customers travels with public transport, and leads to the following theorem. 
\begin{theorem}[Equilibria of the Sequential Game]\label{thm:existence_equilibrium}
Let \cref{assumption:delay_1,assumption:delay_2} hold. Then, the sequential game possesses an equilibrium. 
\end{theorem}
The proof of \cref{thm:existence_equilibrium} provides some insights on the equilibrium itself, which we summarize as follows.
\begin{corollary}\label{corollary:equilibrium_explicit}
Let \cref{assumption:delay_1,assumption:delay_2} hold.
For $j\in\{1,\ldots,N\}$, define $\pureStrategy{j}^\ast$ as: 
\begin{enumerate}
\setlength{\leftmargin}{0pt}
    \item If~$\tilde{c}_j + \valueTime\cdot\delayTime{j}(0)\geq \pureStrategy{0}+\valueTime\cdot\delayTime{0}$, then:~$\pureStrategy{j}^\ast(\pureStrategy{0}) = \tup{0,0}$.

    \item 
    Else, let $p_j^\ast\in[\tilde{c}_j,\pureStrategy{0}+\valueTime\cdot(\delayTime{0}-\delayTime{j}(0))]$ be the unique solution of
    \begin{equation}\label{eq:necessary_sufficient_optimum}
    \begin{aligned}
        -\frac{p_j^\ast-\tilde{c}_j}{\valueTime}
        \derivative{\delayTimeInverse{j}}{z}{\frac{\pureStrategy{0}-p_j^\ast}{\valueTime}+\delayTime{0}}
        + \delayTimeInverse{j}\left(\frac{\pureStrategy{0}-p_j^\ast}{\valueTime}+\delayTime{0}\right)=0.
    \end{aligned}
    \end{equation}
    Then,~$\pureStrategy{j}^\ast(\pureStrategy{0}) =    
        \tup{p^\ast_j(\pureStrategy{0}),\delayTimeInverse{j}\left(\frac{\pureStrategy{0}-p_j^\ast(\pureStrategy{0})}{\valueTime}+\delayTime{0}\right)}.$
\end{enumerate}
Further, let~$\pureStrategy{0}^\ast$ be a minimizer of $\utility{0}(\pureStrategy{0},\pureStrategy{-0}^\ast(\pureStrategy{0}))$.
Then,~$\tup{\pureStrategy{0}^\ast,\pureStrategy{1}^\ast,\ldots,\pureStrategy{N}^\ast}$ is an equilibrium. 
\end{corollary}
If we further assume that delay functions are affine, then we can compute equilibria in closed form. 
\begin{corollary}\label{corollary:equilibrium_linear}
    Let \cref{assumption:delay_1,assumption:delay_2} hold and assume that~$\delayTime{j}$ is affine, i.e.,~$\delayTime{j}(z)=\alpha_j+\beta_jz$,  
    with $\alpha_j,\beta_j\geq 0$, 
    and $\tilde{c}_j+\valueTime\cdot\alpha_j < \valueTime\cdot\delayTime{0}$. 
    Then, 
    \begin{equation}\label{eq:affine_strategy_msp}
    \begin{aligned}
        \pureStrategy{j}^\ast(\pureStrategy{0})
        =
        \Big\langle 
        &\frac{\pureStrategy{0} + \valueTime(\delayTime{0}-\alpha_j)-\tilde{c}_j}{2},
        \frac{\pureStrategy{0} + \valueTime(\delayTime{0}-\alpha_j)-\tilde{c}_j}{2\valueTime\beta_j}
        \Big \rangle 
    \end{aligned}
    \end{equation}
    and 
    \begin{equation}\label{eq:affine_strategy_municipality}
    \begin{aligned}
        \pureStrategy{0}^\ast
        =
        \proj{\frac{1-\frac{k_1}{k_3}-\sum_{j=1}^N\frac{\frac{k_2}{k_3}\varepsilon_j+(\tilde{c}_j +\valueTime(\alpha_j-\delayTime{0}))}{2\valueTime\beta_j}}{\sum_{j=1}^N\frac{1}{\valueTime\beta_j}}}{[0,p_0^\mathrm{max}]},
    \end{aligned}
    \end{equation}
    where $\proj{x}{A}$ denotes the projection of $x$ in the set $A$. 
\end{corollary}
\cref{corollary:equilibrium_linear} shows intuitive features in extreme cases: 
\begin{itemize}
    \item A municipality prioritizing the cost for customers (i.e., $k_1=1, k_2,k_3\to 0$) will choose $\pureStrategy{0}^\ast\to 0$. 
    \item A municipality prioritizing the environmental impact (i.e., $k_2=1, k_1,k_3\to 0$) will choose $\pureStrategy{0}^\ast\to 0$. 
    \item A municipality prioritizing its revenue (i.e., $k_3=1, k_1,k_2\to0$) will choose
    \begin{equation*}
    \pureStrategy{0}^\ast\to
    \proj{\frac{1-\sum_{j=1}^N\tilde{c}_j +\valueTime(\alpha_j-\delayTime{0}))}{\sum_{j=1}^N\frac{1}{\valueTime\beta_j}}}{[0,p_0^\mathrm{max}]}.
    \end{equation*}
\end{itemize}

The generalization of this approach to general congestion games is an active field of research~\cite{marchesi2019leadership}, and we leave its treatment to future work.

\begin{table*}[t]
	\begin{center}
		\begin{scriptsize}
		\begin{tabular}{llcccccclc}
			\toprule
			\textbf{Parameter} & \textbf{Variable}  & \multicolumn{6}{c}{\textbf{Value}} &\textbf{Units}& \textbf{Source}\\
			\midrule
			&& \textbf{SV ICEV} & \textbf{SV Hybrid} &\textbf{SV Electric} & \textbf{AV ICEV} & \textbf{AV Hybrid} &\textbf{AV Electric}\\ \cline{3-8}  \\[-1.0em]
			Vehicle operational cost & $c_\mathrm{v,o}$ &\SI{5.78}{} & \SI{5.86}{}& \SI{5.89}{} & \SI{0.38}{} & \SI{0.45}{} & \SI{0.48}{} & \si{\usd\per\mile } & ~\cite{AAA2019, BoeschBeckerEtAl2018}\\
			Vehicle fixed cost & $c_\mathrm{v,f}$&\SI{19000}{} & \SI{31000}{}& \SI{29000}{} & \SI{89000}{} & \SI{101000}{} & \SI{99000}{} & \si{\usd\per\car } & ~\cite{AAA2019, zardini2020cobis}\\
			Vehicle emissions & $e_\mathrm{v}$ &\SI{0.16}{} & \SI{0.18}{}& \SI{0.13}{} & \SI{0.16}{} & \SI{0.18}{} & \SI{0.13}{} & \si{\kg\per\mile } & ~\cite{AAA2019, zardini2020cobis,jochem2015assessing}\\
			Vehicle life & $l_\mathrm{v}$ &\multicolumn{6}{c}{\SI{186000}{}}&\si{miles}&~\cite{AAA2019}\\
			\midrule
			&&&& \textbf{ES} &\textbf{SB}&& \\ \cline{5-6}  \\[-1.0em]
			\gls{abk:mm} operational cost&$c_\mathrm{m,o}$&&& \SI{0.79} &\SI{1.58}&& &\si{\usd\per\mile}&\cite{zardini2020cobis, SchellongEtAl2019}\\
			\gls{abk:mm} speed &$v_\mathrm{m}$& &&\SI{5.0} &\SI{7.5}&& &\si{mph}&\cite{NATCOMicromobility2018}\\
			\gls{abk:mm} emissions &$e_\mathrm{m}$& &&\SI{0.101} &\SI{0.033}&& &\si{\kg\per\mile}&\cite{zardini2020cobis, SchellongEtAl2019}\\
			\midrule
			&&& \textbf{U-Bahn} &\textbf{S-Bahn} & \textbf{Trams} &\textbf{Buses}\\ \cline{4-7}  \\[-1.0em]
			PT waiting time&$t_\mathrm{w,p}$&& 5.0&5.0&7.0&10.0&&\si{min}&\cite{bvg20}\\
		\bottomrule
	\end{tabular}
	\vspace{0.1cm}
	\caption{Parameters, variables, numbers, and units for the case studies.}
	\label{tab:params_1}
\end{scriptsize}
\end{center}
\end{table*}
\subsubsection{Game-theoretic Model of the Mobility System}
We adapt the game-theoretic model of the mobility system from~\cite{LanzettiSchifferEtAl2021}. In particular, we consider a mobility system with an \gls{abk:amod} operator, an \gls{abk:mm} operator, a taxi company, and public transport. We assume that the \gls{abk:amod} operator changes her prices dynamically, while all other \glspl{abk:msp} and the municipality strategically choose their prices on a longer time horizon. In addition to travelling with \glspl{abk:msp} or public transport, customers can also walk from their origin to their destination. Formally, we model the mobility system as a multigraph~$\graph=\tup{\setOfVertices,\setOfArcs,s,t}$, where~$\setOfVertices$ is the set of vertices,~$\setOfArcs$ is the set of arcs, and~$s\colon\setOfArcs\to\setOfVertices$ and~$t\colon\setOfArcs\to\setOfVertices$ map each arc to its origin and destination vertices, respectively. We define arc subsets~$\setOfArcs_\text{\gls{abk:amod}}$,~$\setOfArcs_\text{\gls{abk:mm}}$,~$\setOfArcs_\text{PT}$,~$\setOfArcs_\text{Taxi}$, and~$\setOfArcs_\text{W}$, each inducing a subgraph. 
We model customers demand by a triple~$\tup{\origin,\destination,\rate}$, where~$\origin$ is the origin of the trip,~$\destination$ is its destination, and~$\rate$ is the rate of customers per unit time, and consider $M$ demands. 
We assume that customers of demand~$i\in\{1,\ldots,M\}$ select their trip via a mobile app, offering them two options. First, the \gls{abk:amod} ride, which takes time $t_{\text{\gls{abk:amod}},i}$ and results from the shortest path from the customer's origin to the customer's destination on~$\graph_\text{\gls{abk:amod}}$. Second, the most convenient option between \gls{abk:mm}, public transport, taxi, and walking. The most convenient option is defined as the one minimizing the sum of ticket prices and monetary value of time, and takes time $t_{\mathrm{alt},i}$ (whereby the time is again computed via shortest path).
For each demand, we assume a linear reaction curve: the rate of customers choosing the \gls{abk:amod} ride decreases linearly with the price of the \gls{abk:amod} ride. The parameters of the linear curve depend on the customers' distribution of value of time; see~\cite{LanzettiSchifferEtAl2021,wadud2017fully}.
We consider a transportation system in steady state and suppose that the \gls{abk:amod} operator offers mobility service on her subgraph $\graph_{\text{\gls{abk:amod}}}$. To prevent imbalances in her fleet, the \gls{abk:amod} operator rebalances her fleet via empty vehicles. Formally, we dictate that the sum of vehicle flows entering a node equals the sum of the vehicle flows exiting it. When taking operational decisions, the \gls{abk:amod} operator has a limited number of \glspl{abk:av} and it is subject to two types of taxes imposed by the municipality: a distance-based tax on all \glspl{abk:av} and an additional distance-based tax on \glspl{abk:av} driving empty, without customers.
Then, the \gls{abk:amod} operator selects prices to maximize profit. Formally, prices result from the quadratic convex optimization problem 
\begin{subequations}
\begin{align}
    \max_{\substack{x_i\in[0,\alpha_i], \\ f_0\in\mathbb{R}_{\geq 0}^{|\setOfArcs_{\text{\gls{abk:amod}}}|}}}
    &\sum_{i=1}^M\left[x_i(m_ix_i+q_i) - \sum_{j\in\setOfArcs_\text{\gls{abk:amod}}} c_jf_{i,j}^\ast x_i + c_{0,j}f_{0,j}\right] \nonumber
    \\
    \text{s.t. }
    & \mat{B}^\top \left(f_{i}^\ast x_i + f_0\right)=0 \label{constraint:conservation} \\
    & \sum_i t_{\mathrm{AMoD},i} x_i + \sum_j t_j f_{0,j} \leq N_\mathrm{fleet}, \label{constraint:fleet}
\end{align}
\end{subequations}
where~$x_i$ denotes the flow of customers served by the \gls{abk:amod} operator and~$f_0$ are rebalancing vehicle flows.
Here,~$m_i<0$ and~$q_i$ are parameters defining the reaction curve, depending on the value of time, $\rate_{i}$, $t_{\text{\gls{abk:amod}},i}$, and $t_{\text{alt},i}$. The parameters~$c_j$ and ~$c_{0,j}$ model operational costs, including taxes, corresponding to arc~$j\in\setOfArcs_{j}$. The vector~$f_j\in\{0,1\}^{|\setOfArcs_{\text{\gls{abk:amod}}}|}$ denotes the shortest path on~$\graph_{\text{\gls{abk:amod}}}$ from~$\origin_i$ to~$\destination_i$. Constraint \eqref{constraint:conservation}, with~$\mat{B}$ being the incidence matrix of $\graph_\text{\gls{abk:amod}}$, ensures vehicle conservation at each node. Constraint \eqref{constraint:fleet} upper-bounds the fleet size. Finally, prices result from combining $x_i$ with the customers' reaction curve.
The profit of other \glspl{abk:msp}, defined as excess of revenue over costs, results from the solution of this optimization problem, via the demand function. Finally, social welfare results from the weighted combination of the cost for the customers, CO\textsubscript{2} emissions, and public revenue. Specifically, the cost for the customers results from the fares and the monetary value of time. 
CO\textsubscript{2} emissions result from the distance driven by \glspl{abk:av}, taxi, and \gls{abk:mm} vehicles; here, we neglect emissions of the public transportation systems, as we consider them as fixed and independent from the system's load, therefore not influencing strategic decisions. Finally, public revenue results from taxes and public transport tickets.

Consistently with the described low-level model of the mobility system, we consider a game among a municipality deciding on public transport prices and taxes (specific values, but also generic taxation strategies) for \gls{abk:amod} vehicles. In line with current public transport prices in many cities, we parametrize fares via a \gls{abk:sdp} in~$P_\mathrm{p,s}\subseteq \mathbb{R}_{\geq 0}$, a \gls{abk:ldp} in~$P_\mathrm{p,l}\subseteq \mathbb{R}_{\geq 0}$, and a cutoff distance in~$D\subseteq \mathbb{R}_{\geq 0}$. We consider two types of taxes: a distance-based tax on \glspl{abk:av} in~$T_\mathrm{m}\subseteq \mathbb{R}_{\geq 0}$ and an additional distance-based tax on empty \glspl{abk:av} in~$T_\mathrm{e}\subseteq \mathbb{R}_{\geq 0}$, resulting from the scaling of the first tax. So, overall, the strategy space of the municipality is~$\pureStrategySet{0}=P_\mathrm{p,s}\times P_\mathrm{p,l}\times D\times T_\mathrm{m}\times T_\mathrm{e}$.
The \gls{abk:amod} operator chooses the propulsion type of \glspl{abk:av} from a set of options~$E$ (e.g., \gls{abk:icev},~\gls{abk:hev}, or \gls{abk:bev}), the automation level from~$A$ (e.g., \gls{abk:sv}, \gls{abk:av}), and the fleet size from~$F\subseteq \mathbb{N}_{\geq0}$. Our framework can be easily extended to capture different degrees of vehicle automation~\cite{zardini2020co,zardini2020control, zardini2020iros20}.
As \gls{abk:amod} applies dynamic pricing, decisions on prices do not happen at the level of our game, but are rather embedded in the low-level model of the mobility  system. The action space of the \gls{abk:amod} operator is~$\actionSet{1}(\pureStrategy{0})=\actionSet{1}=E\times A\times F$.
Hence,~$\pureStrategySet{1}$ consists of all maps from~$\pureStrategySet{0}$ to~$\actionSet{1}$. For instance, the action of the \gls{abk:amod} operator if the municipality played~$\pureStrategy{0}$ is~$\pureStrategy{1}(\pureStrategy{0})=\tup{e,a,n}\in E\times A\times F$.
The \gls{abk:mm} operator decides on prices by choosing base and variable, mileage-depedent, prices from~$P_{\mathrm{m,b}}\times P_\mathrm{m,v}\subseteq \mathbb{R}_{\geq0}^2$. She also chooses the type of vehicles from~$M$ (e.g., \gls{abk:es} or \gls{abk:sb}), giving~$\actionSet{2}(\pureStrategy{0})=\actionSet{2}=P_{\mathrm{m,b}}\times P_\mathrm{m,v}\times M$.
Finally, the taxi company decides on base and variable prices from~$P_{\mathrm{t,b}}\times P_\mathrm{t,v}\subseteq \mathbb{R}_{\geq0}^2$, giving~$\actionSet{3}(\pureStrategy{0})=\actionSet{3}=P_{\mathrm{t,b}}\times P_\mathrm{t,v}$.
In the following, we instantiate the model in a numerical case study.
\section{Results}
\label{sec:results}

\newcommand*\circled[1]{\tikz[baseline=(char.base)]{
    \node[shape=circle,draw,inner sep=0.5pt] (char) {\scriptsize #1};}}

\def\widthPlot{3.8cm}
\def\shiftPlot{2.9cm}
\def\customers{Cost for customers [100k \si{\usd\per\hour}]}
\def\revenue{Revenue [100k \si{\usd\per\hour}]}
\def\emissions{Cost emissions [100k \si{\usd\per\hour}]}
\def\customersMin{0.9}
\def\customersMax{1.45}
\def\customersTick{{0.9,1.0,1.1,1.2,1.3,1.4}}
\def\revenueMin{0}
\def\revenueMax{2.75}
\def\revenueTick{{0,0.5,1.0,1.5,2.0,2.5}}
\def\emissionsMin{0}
\def\emissionsMax{11}
\def\emissionsTick{{0,2,4,6,8,10}}
\def\markerSize{0.4pt}

Our case study bases on a real-world setting of the city center of Berlin, Germany. We derive the road network and its features from OpenStreetMap~\cite{haklay2008openstreetmap}, and import the public transport network, including U-Bahn, S-Bahn, tram, and bus lines, together with its schedules from GTFS~\cite{VBB}. We consider a set of 129,560 real travel requests~\cite{ziemke2019matsim}.
To compute $t_\text{\gls{abk:amod}}$ we use an average waiting time of \SI{3}{\minute}~\cite{newsweek}. To compute $t_\text{alt}$ we use public transport schedules, velocity of \gls{abk:mm} vehicles, and an average walking velocity of $\SI{3.13}{mph}$. 
We account for congestion effects by increasing the nominal travel time of each interested arc by \SI{56}{\percent}, corresponding to congestion levels in the evening peaks in Berlin~\cite{tomtom}.
In line with~\cite{wadud2017fully}, we assume the customers' value of time to be uniformly distributed between \SI{10}{\usd\per\hour} and \SI{17}{\usd\per\hour}.
We report the parameters, such as operational costs and emissions in \cref{tab:params_1}, and the action spaces of the players in \cref{tab:actions}.

\begin{table}[t]
	\begin{center}
		\begin{scriptsize}
		\addtolength{\tabcolsep}{-0.18cm}
		\thickmuskip=0.5mu plus 0.5mu
		\begin{tabular}{lllcc}
			\toprule
			\textbf{Pl.} & \textbf{Parameter}  & \textbf{Actions} &\textbf{Units}& \textbf{Source}\\
			\midrule 
			\multirow{3}{*}{\rotatebox[origin=c]{90}{\gls{abk:msp}1}} 
			&Fleet size&$F=\{0,\SI{1000}{},\ldots, \SI{16000}{}\}$&\si{cars}&\cite{Berlin2018Taxi}\\
			&Engine&$E=\{\text{\gls{abk:icev}},\text{\gls{abk:hev}},\text{\gls{abk:bev}}\}$&-&-\\
			&Automation&$A=\{\text{\gls{abk:sv}}, \text{\gls{abk:av}}\}$&-&-\\
			\midrule
		\multirow{3}{*}{\rotatebox[origin=c]{90}{\gls{abk:msp}2}}&Vehicle type&$M=\{\text{\gls{abk:es}},\text{\gls{abk:sb}}\}$&-&-\\
			&Base price&$P_{\mathrm{m,b}}=\{1.20\}$&\si{\usd}&\cite{Berlin21}\\
			&Variable price&$P_{\mathrm{m,v}}=\{0.72, 0.96, 1.21, 1.45, 1.69\}$&\si{\usd \per \mile}&\cite{Berlin2018Taxi, Berlin21}\\
			\midrule
			\multirow{2}{*}{\rotatebox[origin=c]{90}{\gls{abk:msp}3}}&Base price&$P_{\mathrm{t,b}}=\{4.72\}$&\si{\usd}&\cite{Berlin21bis}\\[+2pt]
			&Variable price&$P_{\mathrm{t,v}}=\{1.17, 1.95, 3.89, 5.84,7.79\}$ &\si{\usd \per \mile}&\cite{Berlin21bis}\\
			\midrule
			\multirow{5}{*}{\rotatebox[origin=c]{90}{City}}&\gls{abk:sdp}&$P_\mathrm{p,s}=\{0,1.0,2.0,3.0,4.0\}$&\si{\usd}&\cite{Berlin21}\\
			&\gls{abk:ldp}&$P_\mathrm{p,l}=\{0,1.0,2.0,3.0,4.0,5.0\}$&\si{\usd}&\cite{Berlin21,bvg20}\\
			&Cutoff distance&$D=\{0,1.55,3.10\}$&\si{miles}&\cite{Berlin21}\\
			&Miles tax&$T_\mathrm{m}=\{0, 0.16, 0.32, \ldots, 1.60\}$&\si{\usd \per \mile}&-\\
			&Empty miles tax&$T_\mathrm{e}=T_\mathrm{m}\times \{0,1,10\}$&\si{\usd \per \mile}&-\\
		\bottomrule
	\end{tabular}
	\vspace{0.1cm}
	\caption{Parameters, actions, units, and sources for the case study.}
	\label{tab:actions}
\end{scriptsize}
\end{center}
\end{table}

\begin{figure}[t]
    \begin{center}
        \begin{tikzpicture}[scale=1.0]
    \begin{axis}[width=6cm,
                 height=6cm,
                 label style={font=\footnotesize},
                 tick label style={font=\footnotesize},
                 z label style={yshift=-0.2cm,xshift=-0.4cm},
                 view={-150}{20},
                 xlabel={\customers},
                 ylabel={\emissions},
                 zlabel={\revenue},
                 xmin=\customersMin,
                 xmax=\customersMax,
                 xtick=\customersTick,
                 zmin=\revenueMin,
                 zmax=\revenueMax,
                 ztick=\revenueTick,
                 ymin=\emissionsMin,
                 ymax=\emissionsMax,
                 ytick=\emissionsTick,
                 ]
                 
        \addplot3[color=blue,only marks,mark size=\markerSize] table[x index=0,y index=1,z index=2,col sep=comma] {data/NE.csv};
        
        \coordinate (ne1) at (axis cs:1.3408,3.1544,2.1320);
        \coordinate (ne2) at (axis cs:0.9194,0.0471,0.0000);
        \coordinate (ne3) at (axis cs:0.9116,2.9109,0.0000);
        
    \end{axis}
    
    \coordinate (l1) at (1.4,3.9); 
    \coordinate (l2) at (4.5,3.5); 
    \coordinate (l3) at (4.5,0.8); 
    
    \tikzstyle{mybox} = [rectangle,rounded corners,draw,align=center,font=\tiny,fill=black!10!white,inner sep=1.5pt]
    
    \draw[thick,-] (ne1)--(l1) node[mybox,left]
    {\textbf{Revenue-oriented City}\\
    \num{5000} AVs, ICEV\\
     $\mu$M: ES \\
      Base price $\mu$M \SI{1.20}{\usd} \\
      Var. price $\mu$M \SI{0.96}{\usd\per\mile} \\ 
      SDP \SI{3}{\usd},
      LDP \SI{5}{\usd} \\
      Cutoff \SI{1.55}{\mile} \\
      Tax \SI{1.28}{\usd\per\mile} \\
      Empty Tax \SI{1.28}{\usd\per\mile}};
    
    \draw[thick,-] (ne2)--(l2) node[mybox,right]
    {\textbf{Green City}\\
    \num{0} AVs\\
     $\mu$M: SB \\
     Base price $\mu$M \SI{1.20}{\usd} \\
     Var. price $\mu$M \SI{1.21}{\usd\per\mile} \\
     SDP \SI{0}{\usd},
     LDP \SI{0}{\usd} \\
     Any cutoff distance\\
     Any Tax  \\
     Any Tax };
    
    \draw[thick,-] (ne3)--(l3) node[mybox,right]
    {\textbf{Customers-oriented City}\\
    \num{5000} AVs, ICEV \\
     $\mu$M: ES\\
     Base price $\mu$M \SI{1.20}{\usd}, \\
     Var. price $\mu$M \SI{1.21}{\usd\per\mile} \\ 
     SDP \SI{0}{\usd},
     LDP \SI{0}{\usd} \\
     Tax \SI{0}{\usd\per\mile} \\
     Empty Tax \SI{0}{\usd\per\mile}};
    
    \draw[color=ForestGreen,fill] (ne1) circle (6*\markerSize);
    \draw[color=ForestGreen,fill] (ne2) circle (6*\markerSize);
    \draw[color=ForestGreen,fill] (ne3) circle (6*\markerSize);
   
\end{tikzpicture}
    \end{center}
    \caption{Equilibria of the game with respect to cost for customers, cost of emissions, and public revenue. Each point is a \glspl{abk:ne} of the simultaneous game between \gls{abk:msp}. The equilibrium of the sequential game directly results from the weights of the three metrics in municipality's social welfare.}
    \label{fig:equilibria_3d}
    \vspace{-0.5cm}
\end{figure}
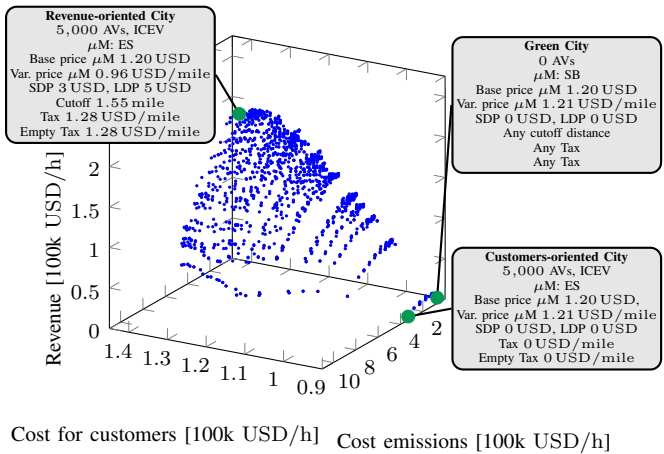

We compute equilibria of the sequential game of \cref{sec:model} via backward induction. First, we look for \glspl{abk:ne} of the simultaneous game between \glspl{abk:msp}\footnote{We report a more exhaustive list of figures at \href{gioele.science/itsc21}{gioele.science/itsc21}.}. We report all of them in \cref{fig:equilibria_3d}, locating them with respect to the three metrics defining social welfare: cost for the customers, emissions, and public revenue. 
Second, we compute the equilibrium of the sequential game by selecting the \gls{abk:ne} maximizing social welfare. As this depends on the weight of each metric, which is a purely political question, we think of these results as drivers to inform policy making.

\begin{figure}[t]
    \begin{center}
        \pgfplotsset{every axis/.append style={
                    axis x line=bottom,
                    axis y line*=left,
                    axis line style={-stealth},
                    label style={font=\tiny},
                    x label style={
                        at={(axis description cs:0.5,-0.1)},
                        anchor=north,
                    },
                    y label style={
                        at={(axis description cs:-0.12,.5)},
                        anchor=south,
                    },
                    x tick label style={font=\tiny,yshift=0.02cm},
                    y tick label style={font=\tiny,xshift=0.08cm},
                    width=\widthPlot,
                    height=\widthPlot,
                    tick align=inside,
                    only marks,
                    mark size=\markerSize,
                    }}


\begin{tikzpicture}[scale=1.0]
    \begin{axis}[xlabel={\emissions},
                 ylabel={\revenue},
                 xmin=\emissionsMin,
                 xmax=\emissionsMax,
                 xtick=\emissionsTick,
                 ymin=\revenueMin,
                 ymax=\revenueMax,
                 ytick=\revenueTick,
                 ]
    
    \addplot+[color=blue,mark=*] table[x index=0,y index=1,col sep=comma] {data/NEEmissionsRevenue.csv};
    \addplot+[color=red,mark=*] table[x index=0,y index=1,col sep=comma] {data/NERationalEmissionsRevenue.csv};
    
    \coordinate (nebetter1) at (axis cs:0.5304,1.7975);
    \coordinate (neworse1) at (axis cs:9.4910,0.7958);
    \coordinate (neincomparable1) at (axis cs:0.7809,0.6469);
    
    \end{axis}
    
    \begin{axis}[xlabel={\customers},
                 ylabel={\revenue},
                 xshift=\shiftPlot,
                 xmin=\customersMin,
                 xmax=\customersMax,
                 xtick=\customersTick,
                 ymin=\revenueMin,
                 ymax=\revenueMax,
                 ytick=\revenueTick,
                 ]
    
    \addplot+[color=blue,mark=*] table[x index=0,y index=1,col sep=comma] {data/NECustomersRevenue.csv};
    \addplot+[color=red,mark=*] table[x index=0,y index=1,col sep=comma] {data/NERationalCustomersRevenue.csv};
    
    \coordinate (nebetter2) at (axis cs:1.2621,1.7975);
    \coordinate (neworse2) at (axis cs:1.3066,0.7958);
    \coordinate (neincomparable2) at (axis cs:1.0219,0.6469);
    
    \end{axis}
    
    \begin{axis}[xlabel={\emissions},
                 ylabel={\customers},
                 xshift=2*\shiftPlot,
                 xmin=\emissionsMin,
                 xmax=\emissionsMax,
                 xtick=\emissionsTick,
                 ymin=\customersMin,
                 ymax=\customersMax,
                 ytick=\customersTick,
                 ]
    
    \addplot+[color=blue,mark=*] table[x index=0,y index=1,col sep=comma] {data/NEEmissionsCustomers.csv};
    \addplot+[color=red,mark=*] table[x index=0,y index=1,col sep=comma] {data/NERationalEmissionsCustomers.csv};
    
    \coordinate (nebetter3) at (axis cs:0.5304,1.2621);
    \coordinate (neworse3) at (axis cs:9.4910,1.3066);
    \coordinate (neincomparable3) at (axis cs:0.7809,1.0219);
    
    \end{axis}
    
    \node[outer sep=0pt,inner sep=0pt] at (4.5,2.1) (one) {\circled{1}};
    \node[outer sep=0pt,inner sep=0pt] at (3.5,1.5) (two) {\circled{2}};
    \node[outer sep=0pt,inner sep=0pt] at (4.0,0.2) (three) {\circled{3}};
    
    \draw[-] (nebetter1) -- (one);
    \draw[-] (nebetter2) -- (one);
    \draw[-] (nebetter3) -- (one);
    
    \draw[-] (neincomparable1) -- (two);
    \draw[-] (neincomparable2) -- (two);
    \draw[-] (neincomparable3) -- (two);
    
    \draw[-] (neworse1) -- (three);
    \draw[-] (neworse2) -- (three);
    \draw[-] (neworse3) -- (three);

    \tikzstyle{better} = [circle,draw,inner sep=4*\markerSize,color=red,fill=red]
    \tikzstyle{worse} = [circle,draw,inner sep=4*\markerSize,color=blue,fill=blue]
    \tikzstyle{incomparable} = [circle,draw,inner sep=4*\markerSize,color=red,fill=red]
    
    \node[better] at (nebetter1) {};
    \node[better] at (nebetter2) {};
    \node[better] at (nebetter3) {};
    
    \node[worse] at (neworse1) {};
    \node[worse] at (neworse2) {};
    \node[worse] at (neworse3) {};
    
    \node[incomparable] at (neincomparable1) {};
    \node[incomparable] at (neincomparable2) {};
    \node[incomparable] at (neincomparable3) {};
    
\end{tikzpicture}
    \end{center}
    \caption{Two-dimensional projections of the equilibria reported in \cref{fig:equilibria_3d}. In red, the \emph{rational} equilibria, dominating the blue ones in the projections.}
    \label{fig:equilibria}
    \vspace{-0.5cm}
\end{figure}
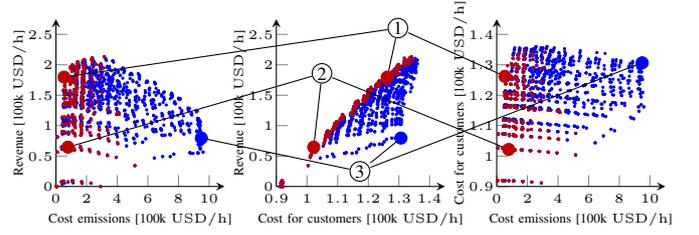

Ideally, from the perspective of a socially-aware municipality, it is desirable to obtain high revenue, low emissions, and low cost for customers. However, \cref{fig:equilibria_3d,fig:equilibria} show that the trade-offs characterizing the game do not allow for such scenarios.
For instance, lowest emissions are reachable with no public revenue and a cost for customers \SI{0.9}{\percent} larger than the potentially obtainable one. Similarly, cost for customers and public revenue are monotonically related. 
Therefore, one needs to come to terms with the fundamental tradeoffs characterizing this system. Nevertheless, our framework provides a way to rigorously reason about possible solutions.

First, fixed the weights of each metric, we provide the equilibrium strategy and the corresponding social welfare. For instance, as can be seen in \cref{fig:equilibria_3d}, a municipality minimizing CO\textsubscript{2} emissions incurs in emissions cost of \SI{4711}{\usd\per\hour}, and should make public transport free and ban \gls{abk:amod} vehicles. A customer-centric city also opts for free public transport and does not introduce taxes, at the price of no public revenue. Conversely, a city maximizing public revenue should make public transport expensive, but not as expensive as possible, and impose a tax of \SI{1.28}{\usd\per\mile} of \gls{abk:amod} vehicles, with an additional tax of \SI{1.28}{\usd\per\mile} on empty \gls{abk:amod} vehicles. These values represent the sweet spot between free public transport and no taxes and too expensive public transport and unsustainable taxes (which also lead to low public revenue). These are three extreme scenarios; for each choice of weights, we can compute the corresponding equilibrium.

Second, while we cannot a priori give an equilibrium, we can identify solutions which are always inconvenient. In general, \gls{abk:ne} are incomparable: is \gls{abk:ne} \circled{1} better than \gls{abk:ne} \circled{2}? 
It depends on the weights of the metrics in social welfare: \gls{abk:ne} \circled{1} yields larger public revenue, but also larger cost for customers. Hence, we call \gls{abk:ne} \circled{1} and \gls{abk:ne} \circled{2} \emph{incomparable}. However, some equilibria are objectively better than others. For instance, \gls{abk:ne} \circled{1} dominates \gls{abk:ne} \circled{3}, as it outperforms it in all three metrics. 
We call non-dominated equilibria \emph{rational}, and depict them in red in \cref{fig:equilibria}.
Interestingly, all \gls{abk:ne} yielding high emissions are dominated, i.e., never rational.

Third, we can study fundamental tradeoffs of the system. For instance, lowering public revenue from \SI{200000}{\usd\per\hour} to \SI{150000}{\usd\per\hour} (i.e., \SI{25}{\percent} reduction) leads to \SI{50}{\percent} lower emissions and \SI{10}{\percent} lower cost for customers.

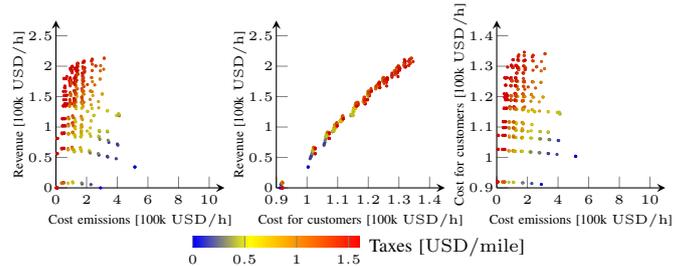
\begin{figure}[t]
    \begin{center}
        \def\colorbarlabel{Taxes [\si{\usd\per\mile}]}
        \def\dataFirst{data/taxEmissionsRevenue.csv}
        \def\dataSecond{data/taxCustomersRevenue.csv}
        \def\dataThird{data/taxEmissionsCustomers.csv}
        \pgfplotsset{every axis/.append style={
                    axis x line=bottom,
                    axis y line*=left,
                    axis line style={-stealth},
                    label style={font=\tiny},
                    x label style={
                        at={(axis description cs:0.5,-0.1)},
                        anchor=north,
                    },
                    y label style={
                        at={(axis description cs:-0.12,.5)},
                        anchor=south,
                    },
                    x tick label style={font=\tiny,yshift=0.02cm},
                    y tick label style={font=\tiny,xshift=0.08cm},
                    width=\widthPlot,
                    height=\widthPlot,
                    tick align=inside,
                    only marks,
                    mark size=\markerSize,
                    scatter,
                    }}

\begin{tikzpicture}[scale=1.0]
    \begin{axis}[xlabel={\emissions},
                 ylabel={\revenue},
                 xmin=\emissionsMin,
                 xmax=\emissionsMax,
                 xtick=\emissionsTick,
                 ymin=\revenueMin,
                 ymax=\revenueMax,
                 ytick=\revenueTick,
                 ]
    
    \addplot+[mark=*] table[x index=0, y index=1, col sep=comma, scatter src=\thisrowno{2}] {\dataFirst};
    
    \end{axis}
    
    \begin{axis}[xlabel={\customers},
                 ylabel={\revenue},
                 xshift=\shiftPlot,
                 xmin=\customersMin,
                 xmax=\customersMax,
                 xtick=\customersTick,
                 ymin=\revenueMin,
                 ymax=\revenueMax,
                 ytick=\revenueTick,
                 colorbar,
                 colorbar horizontal,
                 colorbar style={width=2.2cm,
                                height=0.15cm,
                                axis line style={draw=none},
                                xlabel=\colorbarlabel,
                                x label style={
                                    at={(axis description cs:1.0,0.0)},
                                    anchor=west,
                                    font=\scriptsize,
                                },
                                yshift=0.3cm,
                                xshift=-1.1cm},
                 ]
    
    \addplot+[mark=*,point meta=explicit] table[x index=0, y index=1, col sep=comma, scatter src=\thisrowno{2}] {\dataSecond};
    
    \end{axis}
    
    \begin{axis}[xlabel={\emissions},
                 ylabel={\customers},
                 xshift=2*\shiftPlot,
                 xmin=\emissionsMin,
                 xmax=\emissionsMax,
                 xtick=\emissionsTick,
                 ymin=\customersMin,
                 ymax=\customersMax,
                 ytick=\customersTick,
                 ]
    
    \addplot+[mark=*] table[x index=0, y index=1, col sep=comma, scatter src=\thisrowno{2}] {\dataThird};
    
    \end{axis}
\end{tikzpicture}
    \end{center}
    \caption{Rational equilibria, classified by the entity of the taxes on \gls{abk:amod}.}
    \label{fig:distance_tax}
    \vspace{-0.5cm}
\end{figure}

Fourth, we can ``zoom-in'' and analyze the actions corresponding to each solution, as shown in \cref{fig:distance_tax} for distance-based taxes. As expected, high taxes correlate with high public revenue. However, they also correlate with larger costs for customers, confirming the well-known principle that the tax load partially falls on customers, and not only on sellers.

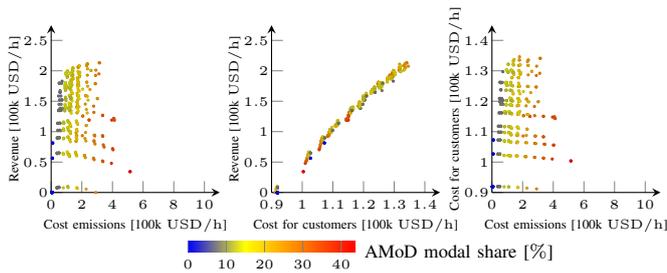
\begin{figure}[t]
    \begin{center}
        \def\colorbarlabel{\gls{abk:amod} modal share [\si{\percent}]}
        \def\dataFirst{data/AMoDShareEmissionsRevenue.csv}
        \def\dataSecond{data/AMoDShareCustomersRevenue.csv}
        \def\dataThird{data/AMoDShareEmissionsCustomers.csv}
        \pgfplotsset{every axis/.append style={
                    axis x line=bottom,
                    axis y line*=left,
                    axis line style={-stealth},
                    label style={font=\tiny},
                    x label style={
                        at={(axis description cs:0.5,-0.1)},
                        anchor=north,
                    },
                    y label style={
                        at={(axis description cs:-0.12,.5)},
                        anchor=south,
                    },
                    x tick label style={font=\tiny,yshift=0.02cm},
                    y tick label style={font=\tiny,xshift=0.08cm},
                    width=\widthPlot,
                    height=\widthPlot,
                    tick align=inside,
                    only marks,
                    mark size=\markerSize,
                    scatter,
                    }}

\begin{tikzpicture}[scale=1.0]
    \begin{axis}[xlabel={\emissions},
                 ylabel={\revenue},
                 xmin=\emissionsMin,
                 xmax=\emissionsMax,
                 xtick=\emissionsTick,
                 ymin=\revenueMin,
                 ymax=\revenueMax,
                 ytick=\revenueTick,
                 ]
    
    \addplot+[mark=*] table[x index=0, y index=1, col sep=comma, scatter src=\thisrowno{2}] {\dataFirst};
    
    \end{axis}
    
    \begin{axis}[xlabel={\customers},
                 ylabel={\revenue},
                 xshift=\shiftPlot,
                 xmin=\customersMin,
                 xmax=\customersMax,
                 xtick=\customersTick,
                 ymin=\revenueMin,
                 ymax=\revenueMax,
                 ytick=\revenueTick,
                 colorbar,
                 colorbar horizontal,
                 colorbar style={width=2.2cm,
                                height=0.15cm,
                                axis line style={draw=none},
                                xlabel=\colorbarlabel,
                                x label style={
                                    at={(axis description cs:1.0,0.0)},
                                    anchor=west,
                                    font=\scriptsize,
                                },
                                yshift=0.3cm,
                                xshift=-1.1cm},
                 ]
    
    \addplot+[mark=*,point meta=explicit] table[x index=0, y index=1, col sep=comma, scatter src=\thisrowno{2}] {\dataSecond};
    
    \end{axis}
    
    \begin{axis}[xlabel={\emissions},
                 ylabel={\customers},
                 xshift=2*\shiftPlot,
                 xmin=\emissionsMin,
                 xmax=\emissionsMax,
                 xtick=\emissionsTick,
                 ymin=\customersMin,
                 ymax=\customersMax,
                 ytick=\customersTick,
                 ]
    
    \addplot+[mark=*] table[x index=0, y index=1, col sep=comma, scatter src=\thisrowno{2}] {\dataThird};
    
    \end{axis}
\end{tikzpicture}
    \end{center}
    \caption{Rational equilibria, classified by the modal share of \gls{abk:amod} vehicles.}
    \label{fig:modal_share_amod}
    \vspace{-0.5cm}
\end{figure}

Fifth, we can evaluate the impact of interventions on other metrics. For instance, \cref{fig:modal_share_amod} shows the \gls{abk:amod} modal share. As expected, higher emissions correlate with larger \gls{abk:amod} modal share. Interestingly, though, we do not observe a correlation with the cost for customers: there are \gls{abk:ne} yielding to low costs and low \gls{abk:amod} modal share as well as \gls{abk:ne} yielding to low costs and large \gls{abk:amod} modal share. 

Sixth, we analyzed results from the perspective of a socially-aware municipality. Yet, our framework can be directly used by (profit-oriented) \glspl{abk:msp} to reason on strategic decisions. 
\section{Conclusion}\label{sec:conclusion}
In this paper, we proposed a game-theoretic framework to study interactions between stakeholders of the mobility ecosystem. Our framework relies on the theory of sequential games, and can modularly accommodate different low-level models of the mobility system. We instantiated our framework in two case studies, a parallel arc congestion game and a game-theoretic model of the transportation system, and study them both analytically and numerically. 
With our framework, we arm stakeholders of the mobility ecosystem with analytical tools to reason about interventions and tradeoffs in mobility systems. 
Our work opens the field for various future research streams. 
First, we would like to instantiate our framework for various classes of low-level models of the mobility system, by explicitly characterizing equilibria and studying algorithms to efficiently compute them. 
Second, we want to exploit our framework to model and study interactions happening at different time scales.
Third, we aim at applying our methodology to study other settings with similar structures, such as energy and global maritime shipping markets.

\bibliographystyle{IEEEtran}
\bibliography{paper}
\appendix
\begin{lemma}\label{lemma:wardrop}
Let \cref{assumption:delay_1,assumption:delay_2} hold. Assume~$f_j\geq 1$ for all~$j\in\{1,\ldots,N\}$.
The Wardrop equilibrium is
\begin{equation*}
\begin{aligned}
    x_j^\ast(p_j,p_0)
    =
    \begin{cases}
        \delayTimeInverse{i}\left(\frac{p_0-p_j}{\valueTime}+\delayTime{0}\right)
        & \text{if } \frac{p_0-p_j}{\valueTime}+\delayTime{0}\geq\delayTime{j}(0), \\
        0 & \text{else}. 
    \end{cases}
\end{aligned}
\end{equation*}
\end{lemma}

\begin{proof}
The proof follows directly from the KKT conditions of the optimization problem \eqref{eq:wardrop_optimization} for $f_j\geq 1$.
\end{proof}

\begin{lemma}\label{lemma:fleet}
    Any non-trivial equilibrium strategy $\pureStrategy{j}^\ast\in\pureStrategySet{j}$, $j\in\{1,\ldots,N\}$, is of the form
    \begin{equation*}
        \pureStrategy{j}^\ast(\pureStrategy{0})=
        \tup{
        p_j(\pureStrategy{0}),x_j^\ast(p_j(\pureStrategy{0}),\pureStrategy{0})
    }
    \text{ for } p_j((\pureStrategy{0})\in\mathbb{R}_{\geq 0}.
    \end{equation*}
\end{lemma}

\begin{proof}
It suffices to observe that \emph{(i)} $p_j < \tilde c_j$ and \emph{(ii)} $f_j>x_j^\ast(p_j,p_0)$ are always suboptimal. 
\end{proof}

\begin{proof}[Proof of \cref{thm:existence_equilibrium}]
First, by \cref{lemma:fleet}, there no loss of generality in assuming \glspl{abk:msp} only decide on prices.
By \cref{lemma:wardrop}, $x_j^\ast$ only depends on $\pureStrategy{j}$ and $\pureStrategy{0}$, so the game reduces to $N$ parallel single leader-single follower Stackelberg games, whereby \glspl{abk:msp} select the action maximizing their profit. Here, we can without loss of generality assume $\actionSet{j}(\pureStrategy{0})=[\tilde{c}_j,\pureStrategy{0}+\valueTime(\delayTime{0}-\delayTime{j}(0))]$; else, the problem is straightforward. 
Then, 
\begin{equation*}
    \utility{j}(\pureStrategy{j},\pureStrategy{0})
    = 
    (\pureStrategy{j}-\tilde{c}_j)\cdot\delayTimeInverse{i}\left(\frac{\pureStrategy{0}-\pureStrategy{j}}{\valueTime}+\delayTime{0}\right).
\end{equation*}
As $\utility{j}$ is smooth, we can compute its second derivative as  
\begin{equation*}
\begin{aligned}
    \frac{\text{d}^2\utility{j}}{\text{d}\pureStrategy{j}^2}
    &=\frac{\pureStrategy{j}-\tilde{c}_j}{\valueTime^2} \secondDerivative{\delayTimeInverse{j}}{z}{\frac{\pureStrategy{0}-\pureStrategy{j}}{\valueTime}+\delayTime{0}}
    -\frac{2}{\valueTime}\derivative{\delayTimeInverse{i}}{z}{\frac{\pureStrategy{0}-\pureStrategy{j}}{\valueTime}+\delayTime{0}}.
\end{aligned}
\end{equation*}
Since $\delayTime{j}$ strictly increasing and convex, its inverse is strictly increasing and concave. Thus,~$\frac{\text{d}^2\utility{j}}{\text{d}\pureStrategy{j}^2}<0$, and $\utility{j}$ is strictly concave. So, its maximizer is unique and continuous in the parameters. In particular, $\pureStrategy{0}\mapsto\pureStrategy{j}^\ast(\pureStrategy{0})$ is well-defined and continuous. 
Hence, $\pureStrategy{0}\mapsto\utility{0}(\pureStrategy{0},\pureStrategy{-0}^\ast(\pureStrategy{0}))$ is a continuous function, being the composition of continuous functions. Since $\pureStrategySet{0}$ is compact, Weierstrass' thereom ensures the existence of a maximizer $\pureStrategy{0}^\ast$. So, $\tup{\pureStrategy{0}^\ast,\pureStrategy{1}^\ast,\ldots,\pureStrategy{N}^\ast}$ is an equilibrium.
\end{proof}

\begin{proof}[Proof of \cref{corollary:equilibrium_explicit}]
The proof follows directly from the proof of \cref{thm:existence_equilibrium}. Indeed, by strict concavity, \eqref{eq:necessary_sufficient_optimum} is a necessary and sufficient condition for optimality. By the intermediate value theorem, \eqref{eq:necessary_sufficient_optimum} admits at least a solution; uniqueness follows again from strict concavity. 
\end{proof}

\begin{proof}[Proof of \cref{corollary:equilibrium_linear}]
With affine delay functions, we have $\delayTimeInverse{j}(z)=(z-\alpha_j)/\beta_j$. Thus, \eqref{eq:necessary_sufficient_optimum} reduces to 
\begin{equation*}
    -\frac{p_j^\ast-\tilde{c}_j}{\valueTime}\frac{1}{\beta_j} + \frac{1}{\beta_j}\left(\frac{\pureStrategy{0}+\valueTime\cdot\delayTime{0}-p_j^\ast}{\valueTime}-\alpha_j\right)=0,
\end{equation*}
which is easily solved to~$p_j^\ast = (\pureStrategy{0}+\valueTime(\delayTime{0}-\alpha_j) - \tilde{c}_j)/2>0$. 
Then, by \cref{corollary:equilibrium_explicit}, we get \eqref{eq:affine_strategy_msp}. To conclude it suffices to observe that social welfare, with $\pureStrategy{-0}=\pureStrategy{-0}^\ast(\pureStrategy{0})$, reduces to 
\begin{equation*}
\begin{aligned}
    \utility{0}(\pureStrategy{0})
    &=
    k_3\pureStrategy{0}
    -k_1(\pureStrategy{0}+\valueTime\delayTime{0})
    -\sum_{j=1}^N(k_2\varepsilon_j+k_3\pureStrategy{0})\eta_j(\pureStrategy{0})
\end{aligned}
\end{equation*}
with~$\eta_j\coloneqq (\pureStrategy{0}+\valueTime(\delayTime{0}-\alpha_j)-\tilde{c}_j)/(2\valueTime\beta_j)$.
This is a strongly concave function, whose maximum lies at \eqref{eq:affine_strategy_municipality}.
\end{proof}

\newpage

\end{document}